\documentclass[12pt]{article}
\usepackage{graphicx,amssymb}
\begin{document}

\begin{center}
{\Large\bf Dilaton Stabilization and Inflation\\ in the D-brane World}\\[12mm]
Hang Bae Kim\\
{\it Particle and Astrophysics Research Center, Department of Physics,
Hanyang University, Seoul 133-791, Korea}\\
{\tt hbkim@hanyang.ac.kr}
\end{center}

\vspace{5mm}

\begin{abstract}
We study the dilaton stabilization in the D-brane world
in which a D-brane constitutes our universe.
The dilaton can be stabilized due to the interplay between the D-brane tension
and the negative scalar curvature of extra dimensions.
Cosmic evolution of the dilaton is investigated with the obtained dilaton potential
and it is found that inflation can be realized before the settlement of the dilaton.
\end{abstract}

\newpage

\section{Introduction}

Dilaton and moduli fields play an important role in string theory.
The vacuum expectation value of the dilaton determines both the gauge
and gravitational coupling constants of the low energy theory.
This is also true for the D-brane world where our universe is supposed
to be a part of a D-brane imbedded on a higher dimensional spacetime.
The D-brane world \cite{D-brane-world} is a realization within string theory
of the brane world idea \cite{brane-world} which attracted much attention in recent years.
It is less ad-hoc than the primitive brane world models
in the sense that we know what dynamical degrees of freedom can be there and
what their actions look like.

The general problem of dilaton and moduli in string theory is that their
potential is flat to all orders in perturbation theory and their vacuum expectation
values cannot be determined. Therefore, some non-perturbative effects
must be involved to stabilize them.
For this purpose, the gaugino condensation, non-perturbative corrections of K\"ahler
potential and various non-trivial form field fluxes were considered previously
\cite{non-perturbative,Choi:1998nx,Lukas:1996zq}.
In this paper, we investigate the dilaton stabilization problem in the D-brane world.
In the D-brane world, the D-brane itself is a non-perturbative setup.
It induces a dilaton potential since the dilaton couples to the D-brane.
But the D-brane alone gives a potential running away toward the weak coupling limit
as the gaugino condensation does. We need a non-perturbative term which drives
the dilaton toward the strong coupling limit.
A good candidate which does this is the form field flux \cite{Lukas:1996zq}.
The D-brane world includes the antisymmetric tensor field $B_{\mu\nu}$
which possibly plays the role.
However, the D-brane is coupled to $B_{\mu\nu}$ and induces a potential for $B_{\mu\nu}$,
which makes the $B_{\mu\nu}$ field massive.
The consequence is that the effect of $B_{\mu\nu}$ becomes time-dependent
and vanishes as the $B_{\mu\nu}$ field settles down to the potential minimum
in the end \cite{Chun:2005ee,cckk}.

Looking for another possibility,
we pay attention to the constant term in the string frame action.
This term can arise from, for example, the non-vanishing scalar curvature of extra dimensions.
It also gives rise to the run-away potential, but its signature can either be positive
or negative depending on the sign of the scalar curvature.
We found that the dilaton can be stabilized due to the interplay between the D-brane tension
and the negative scalar curvature of extra dimensions.
The obtained dilaton potential has a nice feature that it has not only a global minimum
but also flat plateau in the weak coupling region which can be used to realize
the inflationary universe before the settlement of the dilaton in the early universe.

The paper is organized as follows. In section 2, we describe the four-dimensional
effective action of the D-brane world and derive the field equations.
In section 3, the dilaton stabilization in the D-brane world is discussed.
In section 4, the cosmological evolution of the dilaton field is investigated
with the obtained dilaton potential. Section 5 concludes the paper.

\section{Effective Field Theory of the D-brane World}

The D-brane world assumes that our universe is a part of a D$p$-brane
with extra-dimensions are compactified.
Then the bosonic sector consists of the U(1) gauge field $A_\mu$
living on the D$p$-brane and the bulk degrees including the graviton
$g_{\mu\nu}$, the dilaton $\Phi$, and the antisymmetric tensor field
of rank-two $B_{\mu\nu}$.
In the presence of the brane, gauge invariance of $B_{\mu\nu}$ is restored
through its coupling to a U(1) gauge field $A_\mu$ and the gauge invariant
field strength is \cite{Witten:1995im}
\begin{equation}
{\cal B}_{\mu\nu} \equiv B_{\mu\nu}+2\pi\alpha'F_{\mu\nu},
\qquad{\rm where}\qquad F_{\mu\nu}=\partial_\mu A_\nu-\partial_\nu A_\mu.
\end{equation}
The four-dimensional effective action of the bosonic sector in the string frame is
\cite{Chun:2005ee}
\begin{eqnarray}
\label{action-S}
S_{4}&=& \frac{1}{2\kappa_4^2} \int d^4\tilde x \sqrt{-\tilde g}
\left[ e^{-2\Phi}\left(\tilde R-2\bar\Lambda+4\tilde\nabla_\mu\Phi\tilde\nabla^\mu\Phi
-\frac{1}{12} \tilde H_{\mu\nu\rho}\tilde H^{\mu\nu\rho}\right)
\vphantom{\sqrt{\frac12}}\right. \nonumber\\ && \hspace{25mm} \left.
-{m_B^2}e^{-\Phi} \sqrt{1 + \frac12\tilde{\cal B}_{\mu\nu}\tilde{\cal B}^{\mu\nu}
 - \frac{1}{16}\left(\tilde{\cal B}^*_{\mu\nu}\tilde{\cal B}^{\mu\nu}\right)^2}\ \right],
\end{eqnarray}
where the tilde denotes the string frame quantity and
${\cal B}^{\ast}_{\mu\nu}=\frac12
\sqrt{-g}\epsilon_{\mu\nu\alpha\beta} {\cal B}^{\alpha\beta}$
with $\epsilon_{0123} = 1$.
$m_B$ is a parameter defined by $m_B^2=2\kappa_4^2{\cal T}_3$
where ${\cal T}_3$ is the effective brane tension.
If we assume that the six extra-dimensions are compactified with a common radius
$R_c$ and the dilaton is stabilized to give a finite string coupling
$g_s=\langle e^\Phi\rangle$, it is given by
$m_B = \pi^{\frac14}(g_s^2/4\pi)^{\frac{p-3}{16}} \left(R_cM_P\right)^{\frac{15-p}{8}}M_P$
where $M_P=2.4\times10^{18}{\rm GeV}$ is the four-dimensional Planck mass.

The action, and thus also field equations, can be written in a more familiar form
in the Einstein metric, which are defined by
\begin{equation}
g_{\mu\nu} = e^{-2\Phi}\tilde g_{\mu\nu}.
\end{equation}
We will work in this metric from now on.
In terms of Einstein metric, the action becomes
\begin{eqnarray}
\label{action-E}
S_{\rm E}&=& \frac{1}{2\kappa_4^2} \int d^4x \sqrt{-g} \left[
R -2\bar\Lambda e^{2\Phi} -2(\nabla\Phi)^2
-\frac{1}{12}e^{-4\Phi}H^2
\vphantom{\sqrt{\frac12}}\right. \nonumber\\ && \hspace{5mm} \left.
-{m_B^2}e^{3\Phi}\sqrt{1 + \frac12e^{-4\Phi}{\cal B}^2
- \frac{1}{16}e^{-8\Phi}\left({\cal B}^*{\cal B}\right)^2}
-2\Lambda \right].
\end{eqnarray}
The last term is added by hand to incorporate the various contributions to
the cosmological constant.
The origins of $\Lambda$ and $\bar\Lambda$ can be quite varied.
One obvious origin of the $\bar\Lambda$ term is the scalar curvature of extra dimensions
integrated over the whole extra dimensions.
There can be many known or unknown sources for the cosmological constant term $\Lambda$.
In this paper, we will not pursue their origins in detail.
Instead, we treat them as parameters and investigate their consequences.
Manifestly, we use the $\Lambda$ to adjust the four-dimensional effective cosmological
constant to vanish.

The field equations derived from the action are
\begin{equation}
\label{B-eq1}
\nabla_\lambda H_{\lambda\mu\nu} -4H_{\lambda\mu\nu}\nabla^\lambda\Phi
-m_B^2e^{3\Phi}\frac{{\cal B}_{\mu\nu}-\frac14e^{-4\Phi}{\cal B}^*_{\mu\nu}
\left({\cal B}{\cal B}^*\right)}{\sqrt{1+\frac12e^{-4\Phi}{\cal B}^2
-\frac{1}{16}e^{-8\Phi}\left({\cal B}{\cal B}^*\right)^2}} = 0,
\end{equation}
\begin{equation}
\label{P-eq1}
-\nabla^2\Phi + \frac{\partial V(\Phi)}{\partial\Phi} = 0,
\end{equation}
\begin{equation}
\label{E-eq1}
G_{\mu\nu} = \kappa_4^2 T_{\mu\nu},
\end{equation}
where the dilaton potential is
\begin{eqnarray}
V(\Phi) &=& \frac14\left[
2\Lambda + 2\bar\Lambda e^{2\Phi} + \frac{1}{12}e^{-4\Phi}H^2
\vphantom{\sqrt{\frac12}}\right.\nonumber\\&&\hspace{10mm}\left.
{}+m_B^2e^{3\Phi}\sqrt{1 + \frac12e^{-4\Phi}{\cal B}^2
- \frac{1}{16}e^{-8\Phi}\left({\cal B}^*{\cal B}\right)^2}
\ \right],
\label{dilaton-potential}
\end{eqnarray}
and the energy-momentum tensor is
\begin{eqnarray}
\kappa_4^2 T_{\mu\nu}&=& -g_{\mu\nu}\Lambda -g_{\mu\nu}\bar\Lambda e^{2\Phi}
+2\nabla_\mu\Phi\nabla_\nu\Phi-g_{\mu\nu}(\nabla\Phi)^2
\nonumber\\ &&
{}+\frac{1}{12}e^{-4\Phi}
\left(3H_{\mu\lambda\rho}H_\nu^{\;\lambda\rho}
-\frac12g_{\mu\nu}H^2\right)
\nonumber\\ &&
{}+\frac12m_B^2e^{3\Phi} \frac{-g_{\mu\nu}-\frac12g_{\mu\nu}e^{-4\Phi}{\cal B}^2
+e^{-8\Phi}{\cal B}_{\mu\lambda}{\cal B}_\nu^{\;\lambda}}
{\sqrt{1+\frac12e^{-4\Phi}{\cal B}^2
-\frac{1}{16}e^{-8\Phi}\left({\cal B}{\cal B}^*\right)^2}}.
\label{TB}
\end{eqnarray}
We will discuss the dilaton stabilization and the cosmic evolution of the dilaton
based on the potential (\ref{dilaton-potential}) and
the equations (\ref{B-eq1})--(\ref{E-eq1}).

\section{Dilaton Stabilization}

The vacuum expectation value of the dilaton $\Phi$ fixes the string coupling constant by
$\langle e^\Phi\rangle = g_s$,
and determines gravitational couplings of various fields as in Eq.~(\ref{TB}).
Thus the time evolution of the dilaton gives the time-varying gravitational coupling constant.
It can also be attributed to the time-variation of the gauge coupling constants,
and such variations are constrained by the observations of the quasar absorption lines,
the cosmic microwave background, and primordial nucleosynthesis \cite{time-variation}.
Thus the dilaton must be stabilized to the local or global minimum of the potential
at least before nucleosynthesis.

With the potential $V(\Phi)$ in Eq.~(\ref{dilaton-potential}), we study the dilaton
stabilization problem in the D-brane world.
Let us first consider the contribution of the antisymmetric tensor field ${\cal B}_{\mu\nu}$
to the dilaton potential.
The non-vanishing $H$ flux gives the dilaton potential of negative exponential.
It was noticed that the non-vanishing fluxes of various form field flux can give rise to
the dilaton potential and were used to stabilize the dilaton and moduli.
However, for the ${\cal B}_{\mu\nu}$ field in the D-brane world,
it is coupled to the D-brane and
the D-brane brings the DBI-type action for the ${\cal B}$ field.
It acts as the potential for ${\cal B}$ field and derives ${\cal B}$ to zero.
Thus, the flux itself is subject to cosmological evolution and its effect can be
time-dependent and vanish at the end. Furthermore, the form filed flux in general
leads to an anisotropic universe even when it is homogeneous \cite{copeland:1995}.
We will treat the dynamics of non-vanishing ${\cal B}$ filed elsewhere \cite{cckk}
and focus on the case of the vanishing ${\cal B}$ field in this paper.
When ${\cal B}$ vanishes, the effect of the brane on the dilaton potential is
due to the brane tension and it leaves the positive exponential term.
Thus, the dilaton potential we are considering is
\begin{equation}
\label{dilaton-potential-2}
V(\Phi) = \frac12 \left(\Lambda + \bar\Lambda e^{2\Phi} + \frac12m_B^2e^{3\Phi} \right),
\end{equation}
where we included the $\Lambda$ term to make the effective cosmological constant
vanish.

There may be additional non-perturbative contributions to the dilaton potential.
For example, a single gaugino condensation gives rise to the potential of the form
$V_{\rm np}(\Phi) = \mu^2 e^{-\beta e^{-\Phi}}$,
where $\beta$ is a constant which depends on the gauge group
and $\mu$ is related to the supersymmetry breaking scale \cite{non-perturbative}.
The various form field fluxes appearing in string theory can also give rise to
the dilaton potential of the form
$V_{\rm np} = \mu^2 e^{-\gamma\Phi}$ \cite{Lukas:1996zq}.
We assume that they are absent or relatively small compared to the terms
in the potential (\ref{dilaton-potential-2}).

The last term in the potential (\ref{dilaton-potential-2}) is due to the D-brane tension
and an exponentially increasing function of $\Phi$.
It derives the dilaton to run away to the weak coupling limit ($\Phi\rightarrow-\infty$).
As is mentioned above, the non-vanishing form field flux, if it exists,
gives the dilaton potential having an opposite behavior.
This applies also to the antisymmetric tensor field ${\cal B}$,
as explicitly shown in (\ref{dilaton-potential}).
If both types of terms exist together, the dilaton potential has a global minimum
and the stabilization of the dilaton is guaranteed.
But the ${\cal B}$ field in the D-brane world rolls down to ${\cal B}=0$
due to the effect of brane, and its effect on the dilaton is transient.
Sometimes, two or more terms of the same type of exponential can contrive
to have a local minimum, as in the racetrack models or the coalescence of several fluxes.
But, it is known that such models usually suffer from the overshoot problem coming from
the shallowness of the local minimum.

\begin{figure}
\begin{center}
\includegraphics[width=0.5\textwidth]{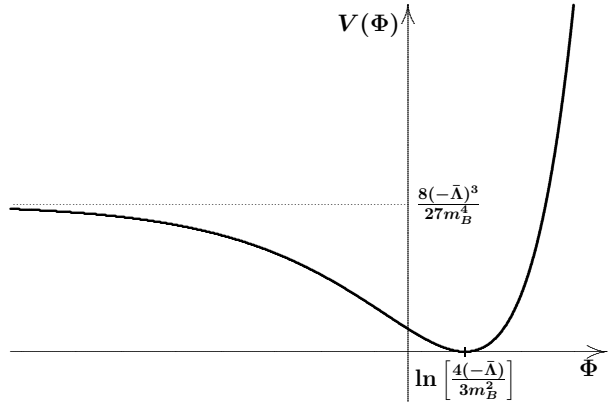}
\end{center}
\caption{The shape of the dilaton potential (\ref{dilaton-potential-2}) when $\bar\Lambda<0$.}
\end{figure}

Without additional non-perturbative terms,
the potential (\ref{dilaton-potential-2}) has positive exponential terms only.
But, there is an interesting possibility to achieve the dilaton stabilization.
The signature of $\bar\Lambda$ can be either positive or negative
depending on underlying theory.
One possible origin of $\bar\Lambda$ is the scalar curvature of extra dimensions
integrated over the extra dimensions themselves.
Therefore, for extra dimensions having negative scalar curvature, $\bar\Lambda$ is negative.
When $\bar\Lambda<0$, the potential has the minimum at
\begin{equation}
\label{minimum}
e^{\Phi_0} = \frac{4(-\bar\Lambda)}{3m_B^2}.
\end{equation}
Thus, depending on relative scales of $-\bar\Lambda$ and $m_B^2$,
we may get the proper vacuum expectation value of the dilaton.
Natural expectation is that they are comparable, and we get the string coupling of order 1.
To make the cosmological constant vanish, $\Lambda$ is adjusted to
\begin{equation}
\Lambda=\frac{8}{27}\frac{(-\bar\Lambda)^3}{m_B^2}.
\end{equation}
The shape of the potential is depicted in Figure~1.
The dilaton mass at the minimum is given by
\begin{equation}
m_\Phi^2=V''(\Phi_0)=\frac{16}{9}\frac{(-\bar\Lambda)^3}{m_B^4}.
\end{equation}
The minimum (\ref{minimum}) is the global one and the final settlement to the minimum
during cosmic evolution is guaranteed if the cosmological constant vanishes.
But its cosmological consequences can be varied depending on how the dilaton evolves.

\section{Cosmological Evolution}

In this section, we investigate the cosmic evolution of the dilaton field
under the potential (\ref{dilaton-potential-2}).
We assume a spatially homogeneous
configuration for the dilaton, and look for the time evolution of the dilaton
and the expansion of the universe.
The existence of non-zero antisymmetric tensor field ${\cal B}$,
even when they are homogeneous, leads to an anisotropic universe in general.
It give rise to a potential for the dilaton
and thus will affect the evolution of the dilaton field.
But in this paper, we consider the case of vanishing antisymmetric tensor field.
The effect of non-vanishing ${\cal B}$ field will be treated elsewhere.
Because the minimum of the potential is the global minimum
and the expansion of the universe induces the friction,
the dilaton will be settled down to it irrespective of the initial conditions.

We assume the metric to be the flat Robertson-Walker metric
\begin{equation}
ds^2 = -dt^2 + a(t)^2\left[(dx^1)^2+(dx^2)^2+(dx^3)^2\right].
\end{equation}
Then the dilaton equation is
\begin{equation}
\label{P-eq3}
\ddot\Phi + 3\dot\alpha\dot\Phi + \frac{\partial V(\Phi)}{\partial\Phi} = 0,
\end{equation}
where $V(\Phi)$ is the dilaton potential in (\ref{dilaton-potential-2})
and we defined $\alpha\equiv\ln a$.

\paragraph{Dilaton-dominated case}
When the dilaton dominates the density of the universe, the Einstein equations are
\begin{eqnarray}
3\dot\alpha^2 &=& \dot\Phi^2 + 2V(\Phi) \\
\ddot\alpha+3\dot\alpha^2 &=& 2V(\Phi).
\label{E-eq3}
\end{eqnarray}
We can solve the combined equations (\ref{P-eq3}) and (\ref{E-eq3}) numerically,
with appropriate initial conditions.
The dimensionless time variable is $m_Bt$ and the mass scale parameter of the potential
is $(-\bar\Lambda)/m_B^2$.
Because the potential is nearly flat in the left and very stiff in the right
of the minimum, the cosmic evolutions are quite distinguished depending on
whether the initial value of the dilaton $\Phi_i$ is smaller or larger than $\Phi_0$.
In Fugure~2, we show the cosmic evolutions of the dilaton field and the scale factor
for typical initial values $\Phi_i$ with $\Phi_i>\Phi_0$ and $\Phi_i<\Phi_0$
for the case of $(-\bar\Lambda)/m_B^2=1$.

\begin{figure}
\begin{center}
\parbox{0.5\textwidth}{
\includegraphics[width=0.45\textwidth]{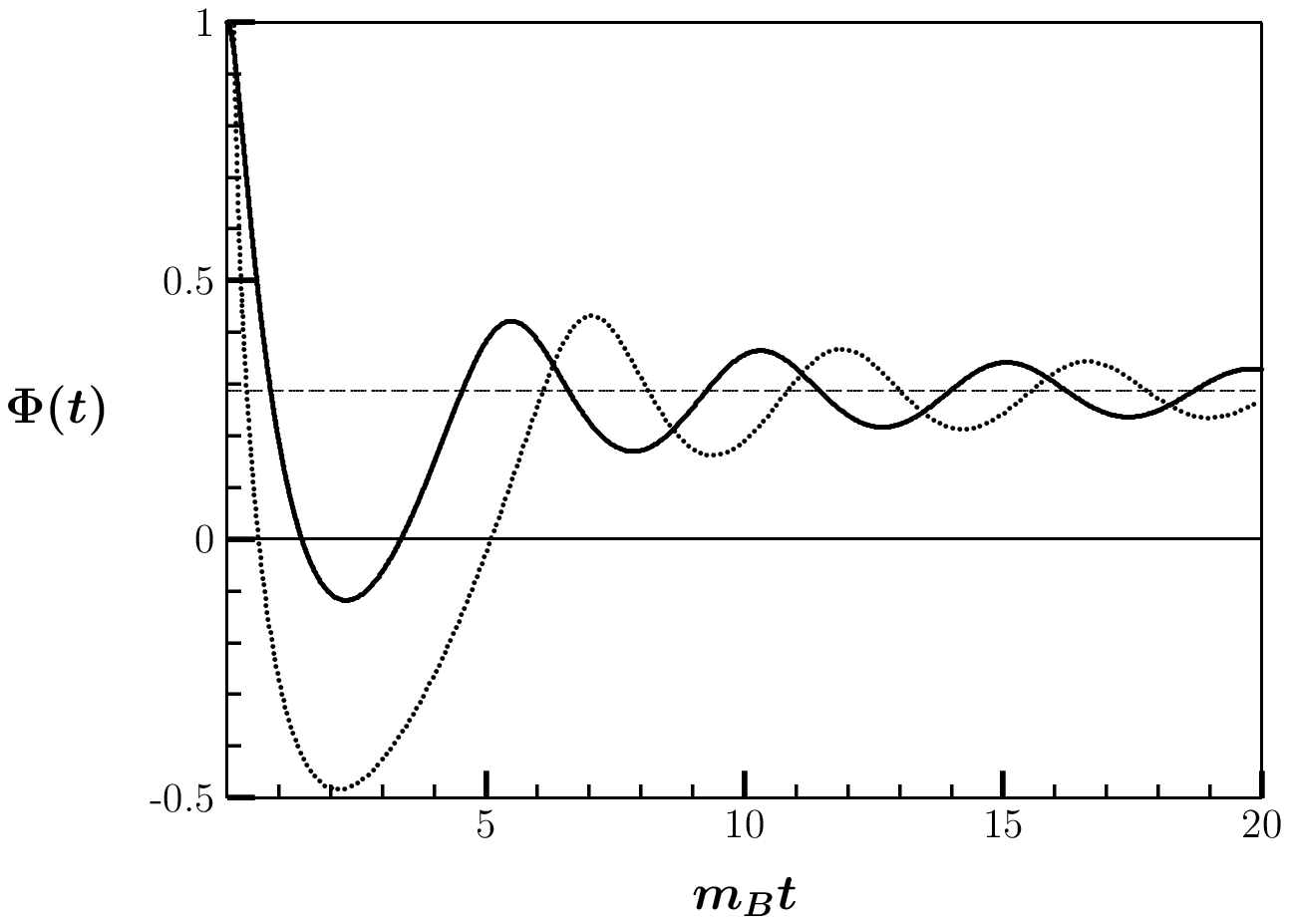}\\
\includegraphics[width=0.45\textwidth]{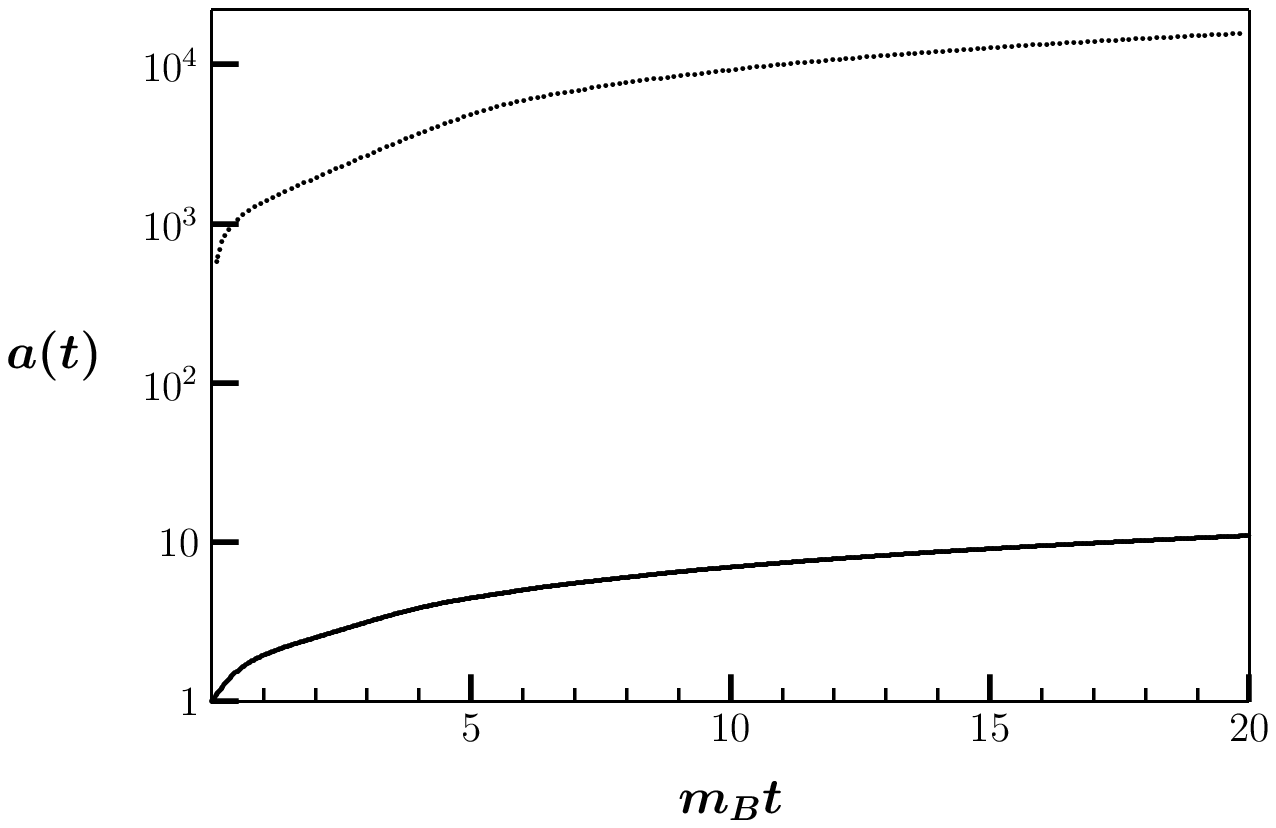}\\
\includegraphics[width=0.45\textwidth]{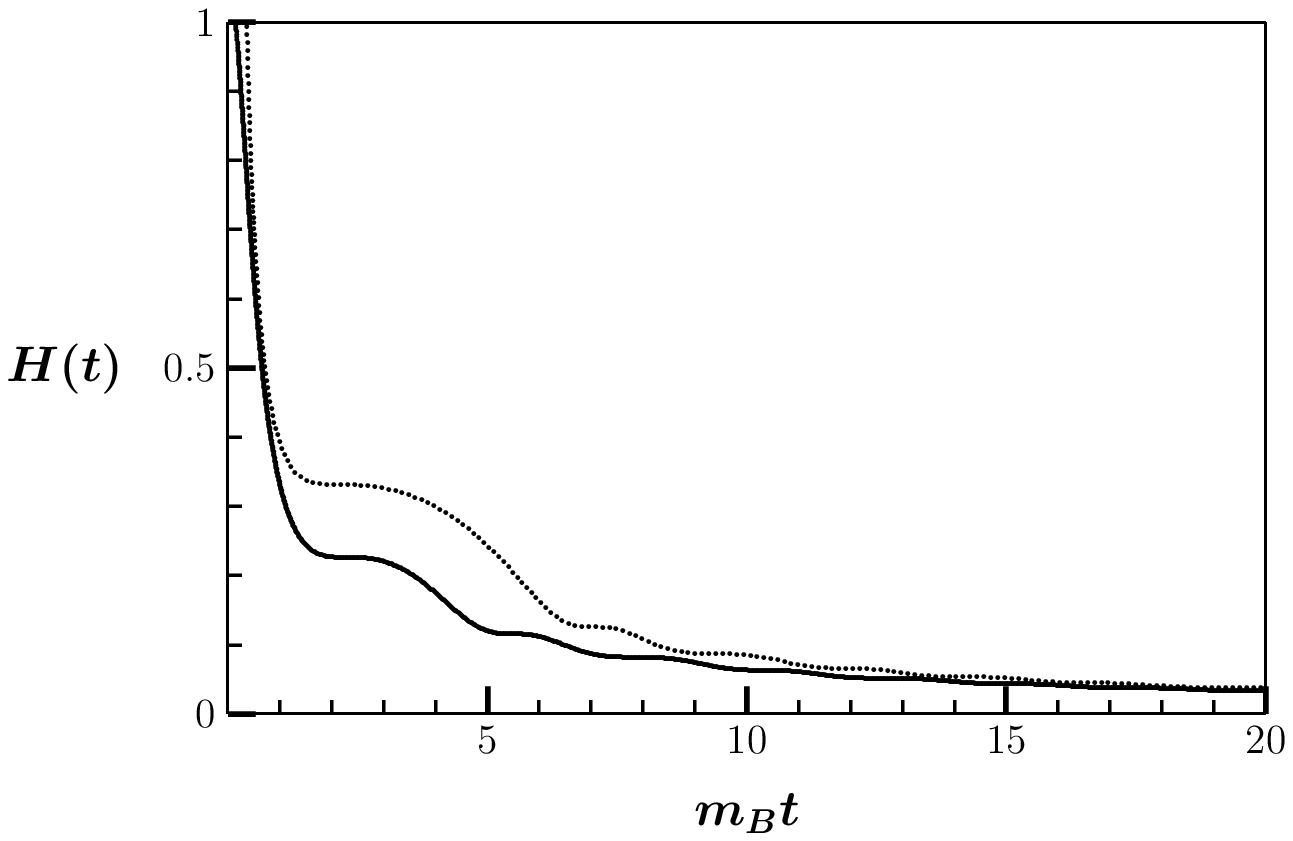}}~%
\parbox{0.5\textwidth}{
\includegraphics[width=0.45\textwidth]{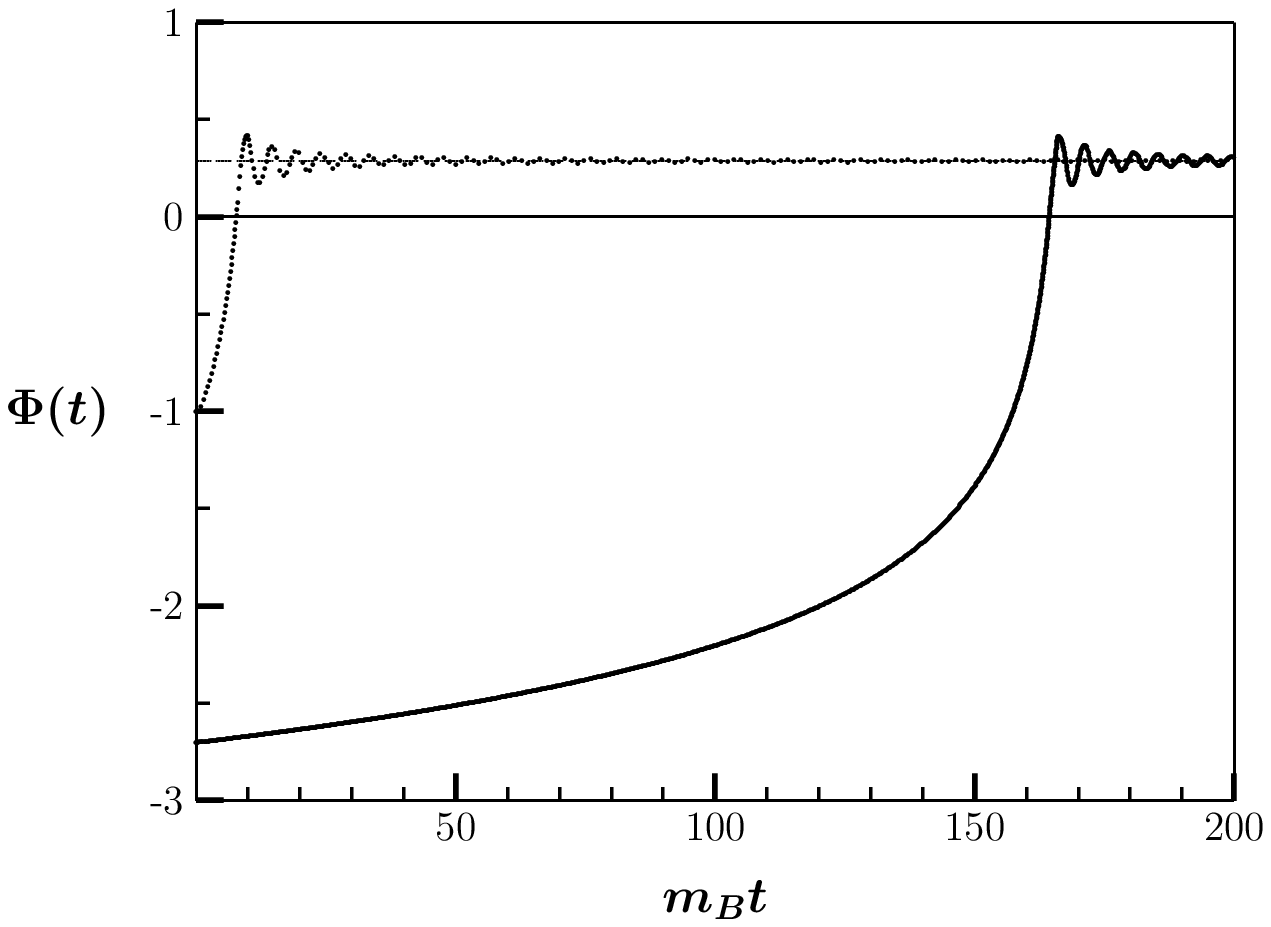}\\
\includegraphics[width=0.45\textwidth]{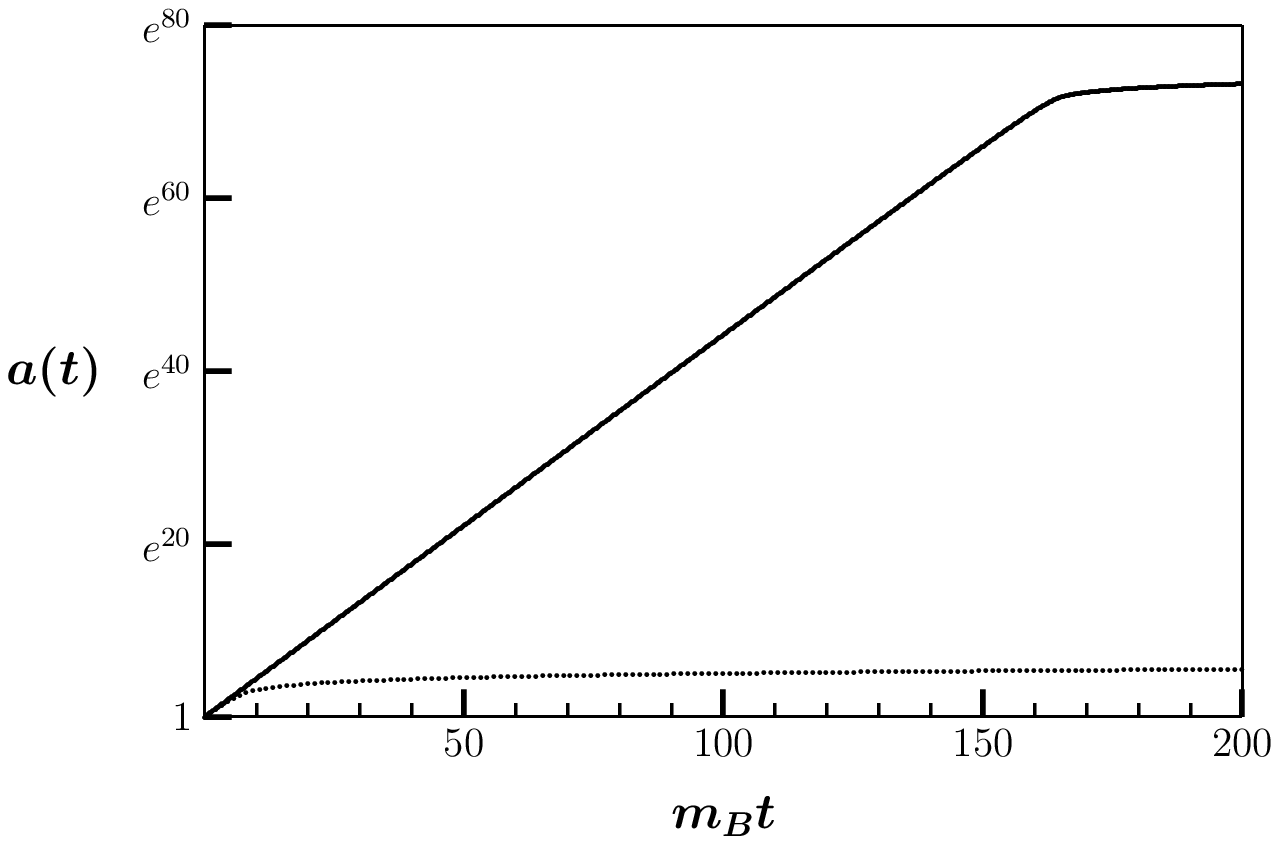}\\
\includegraphics[width=0.45\textwidth]{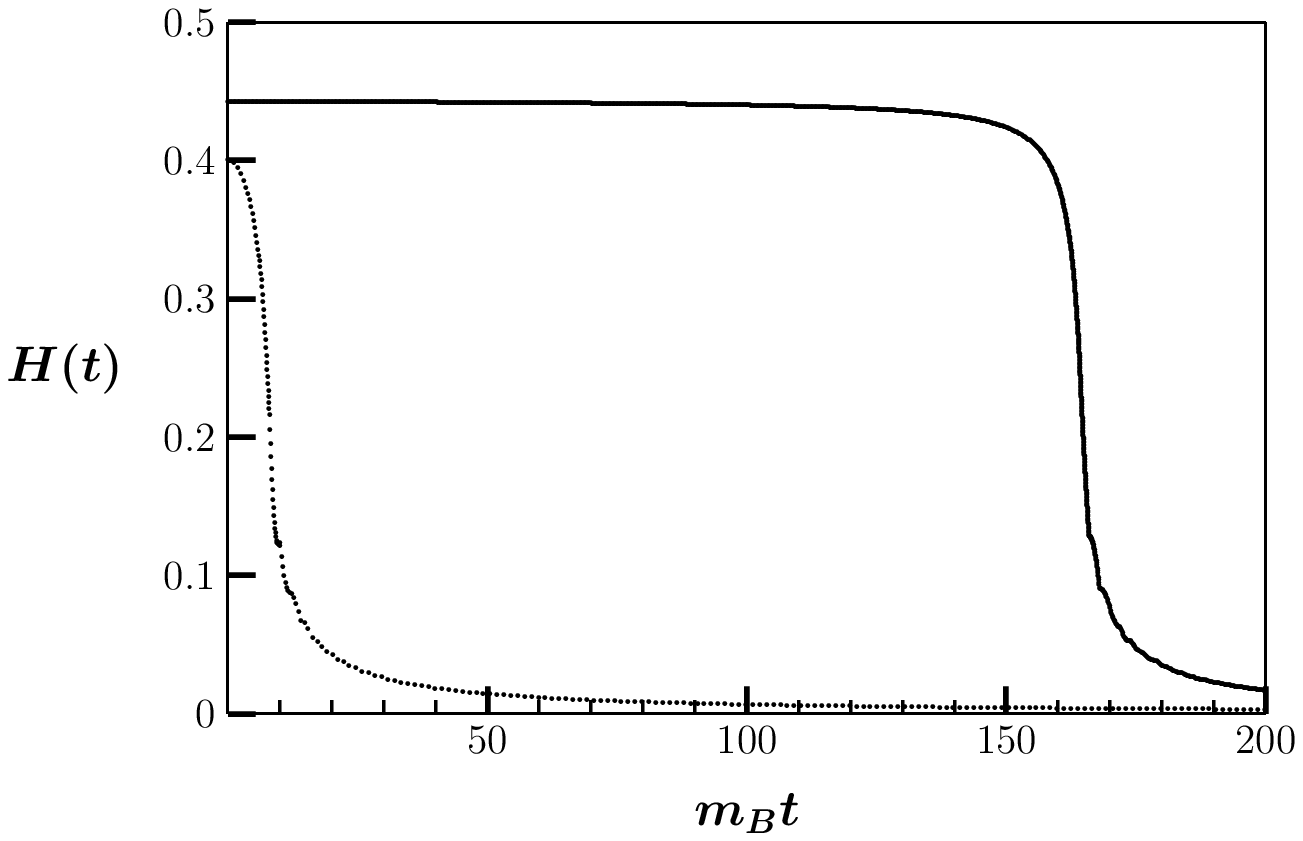}}
\end{center}
\caption{The time evolution of the dilaton $\Phi(t)$ and the scale factor $a(t)$
for $(-\bar\Lambda)/m_B^2=1$. Left Pannels: $\Phi_i>\Phi_0=\ln(4/3)$ ($\Phi_i=1$ for the solid line
and $\Phi_i=10$ for the dotted line). Right Pannels: $\Phi_i<\Phi_0$
($\Phi_i=-2.7$ for the solid line and $\Phi_i=-1$ for the dotted line).}
\end{figure}

When $\Phi_i>\Phi_0$ (strong coupling region),
$\Phi$ rapidly approaches to $\Phi_0$, within a few $m_B^{-1}$,
irrespective of the size of $\Phi_i$.
Then $\Phi$ oscillate about $\Phi_0$ forming `dilaton matter' as it behaves like matter.
Although there can be a big difference in the initial growth of the scale factor,
the late-time evolution is almost independent of the initial value.
The potential is almost flat for $\Phi\ll\Phi_0$, but
there is no overshoot even for the initial value $\Phi_i$ much larger than $\Phi_0$.
This is a characteristic feature of exponential potential.

When $\Phi_i<\Phi_0$ (weak coupling region), the dilaton $\Phi$ slowly rolls down
the flat part of the potential and finally oscillates around the minimum.
Thus, we have slow-roll inflation ended by dilaton oscillation which forms dilaton matter.
This provides a standard picture of slow-roll inflation followed by inflaton oscillation.
In Figure~2, we showed an example that more than 60 e-foldings are obtained in this way.
The decay of dilaton matter heats the universe and the standard big bang universe follows.
However, we note that to keep the flatness of the dilaton potential in the weak coupling
region, it is important to suppress the form field flux which induces the negative
exponential potential for the dilaton.

\paragraph{Evolution of $\Phi(t)$ in the RW background}
When the universe is dominated by homogeneous and isotropic matter or radiation,
while the dilaton contributes a minor fraction, the dilaton evolution is still
given by (\ref{P-eq3}) with $\dot\alpha(t)$ taken from the background geometry.
When the initial value of $\Phi$ resides near the minimum point where the potential
can be approximated by $V(\Phi)\approx \frac12m_\Phi^2(\Phi-\Phi_0)^2$,
the general reasoning applies here.
During $H\gtrsim m_\Phi$, $\Phi$ is frozen at the initial value.
After $H$ becomes smaller than $m_\Phi$, $\Phi$ oscillates about $\Phi_0$,
forming dilaton matter. The dilatons decay finally and dump some amount of entropy
to the thermal bath of the universe.

\section{Conclusion}

We investigated the dilaton stabilization problem in the D-brane world.
In the D-brane world, the effect of the brane tension combined with that of
negative scalar curvature of extra dimensions can provide a stabilizing potential
for the dilaton. The resulting string coupling depends on the ratio of two mass scales
related to the brane tension and the scalar curvature. When two mass scales are comparable,
we get the string coupling of order 1.
We also examined the cosmic evolution of the dilaton with the obtained potential.
Generically, it leaves `dilaton matter' which is a coherent oscillation of the dilaton field
about the minimum point. It dumps a certain amount of entropy to the thermal bath in the
early universe, whose cosmological impact in general depends on the mass and the lifetime
of the dilaton.
The dilaton potential has the flat plateau in the weak coupling region,
and it can lead to the standard inflationary scenario ending with the dilaton oscillation
and finally producing hot big bang universe through dilaton decays.
To keep the potential flat, it is necessary to suppress the form field flux which induces
the strong potential in the weak coupling region.

\section*{Acknowledgements}

This work was supported by the grant KRF-2002-070-C00022.
The author thank Inyoung Cho, Eung Jin Chun, and Yoonbai Kim for helpful discussions.

\end{document}